\documentclass[12pt]{article}
\usepackage[pctex32]{graphics}
\usepackage{amsmath,amsthm}
\usepackage{amssymb,latexsym}
\usepackage[mathscr]{eucal}
\usepackage{setspace}
\begin{document}

\begin{center}\normalsize\textbf{
Magnetohydrodynamic Turbulence: Generalized Formulation}
\end{center}

\bigskip

\begin{center}Bhimsen K. Shivamoggi\\
University of Central Florida\\
Orlando, FL 32816-1364
\end{center} 

\textbf{Abstract}
A general framework that incorporates the Iroshnikov-Kraichnan (IK) and Goldreich-Sridhar (GS) phenomenalogies of magnetohydrodynamic (MHD) turbulence is developed. This affords a clarification of the regime of validity of the IK model and hence help resolve some controversies in this problem. This general formulation appears to have a certain robustness with respect to the inclusion of compressible effects.

\vspace{.8in}

Magnetohydrodynamic (MHD) flows that occur naturally (like astrophysical situations) and in modern technological systems (like fusion reactors) show turbulence. Early theoretical investigations of MHD turbulence considered the isotropic case. On the latter premise, Iroshnikov ~\cite{Iro} and Kraichnan ~\cite{Kra} (IK) made arguments \'{a} la Kolmogorov ~\cite{Kol1} and proposed that statistical properties of the small-scale components of the velocity and magnetic fields

\medskip
*are controlled by the shear Alfv\'{e}n wave dynamics;

*show, in the limit of large viscous and magnetic Reynolds numbers, some 

\ universality in the inertial range;

\medskip
\noindent
and gave for the total energy spectral density $E(k)$, the behavior $E(k)\sim k^{-\frac{3}{2}}$. Montgomery et al. \cite{Mon} and \cite{She}, Goldreich and Sridhar \cite{Sri} - \cite{Gol2} (GS) pointed out that the isotropy assumption in the IK theory is not a very sound one in the MHD case, thanks to the magnetic field of the large-scale eddies, and the GS theory \cite{Gol2} gave for the energy spectrum in the plane transverse to the magnetic field the behavior $E(k_\bot)\sim k_\bot^{-\frac{5}{3}}$. However, EDQNM closure calculations of two-dimensional (2D) MHD turbulence dominated by non-local interactions in the presence of a mean magnetic field (Pouquet et al. \cite{Pou}) and DNS of MHD turbulence in an applied magnetic field (Maron and Goldreich \cite{Mar}, Muller et al. \cite{Mul} and \cite{Mul2}) showed that the transverse energy spectrum is close to the IK theory $E(k_\bot)\sim k_\bot^{-\frac{3}{2}}$. On the other hand, the 3D DNS (Muller et al. \cite{Mul3}, Haugen et al. \cite{Han}) of MHD turbulence, numerical calculations of 2D MHD turbulence dominated by local interactions (Fyfe et al. \cite{Fyf}) and solar wind measurements (Leamon et al. \cite{Lea}, Goldstein et al. \cite{Gol3}) confirm the GS spectrum $E(k_\perp)\sim k_\perp^{-\frac{5}{3}}$. A resolution of this apparent conflict (Boldyrev [\cite{Bol}) requires clarification of the regime of validity of the IK model which is the objective of this paper. A general framework that incorporates IK and GS phenomenalogies is developed to accomplish this objective.
The IK and GS hypotheses can be shown to follow from the formal analogy between the hydrodynamic and MHD spectral energy density expressions. This general formulation is extended further to include compressibility effects.

Let us write the spectral energy density E(k) as

\begin{equation}
E(k) \sim \frac{V}{k^{2}\tau}
\end{equation}

\noindent
V being the characteristic velocity of the spectral element k. The hydrodynamic eddy turn-over time $\tau$ given by

\begin{equation}
\tau \sim \frac{1}{kV}
\end{equation}

\noindent
then becomes

\begin{equation}
\tau \sim \frac{1}{k^{\frac{3}{2}}E^{\frac{1}{2}}}.
\end{equation}

\noindent
(2) implies that the energy transfer in the hydrodynamic case is \emph{local} in the spectral space which reflects the fact that a large-scale velocity field can be transformed away via Galilean invariance.

If we use the relation 

\begin{equation}
\tau \sim \frac{Ek}{\varepsilon}
\end{equation}

\noindent
$\varepsilon$ being the mean energy transfer rate, (3) leads to the Kolmogorov \cite{Kol} spectrum

\begin{equation}
E(k) \sim \varepsilon^{\frac{2}{3}}k^{-\frac{5}{3}}.
\end{equation}

One may write for the MHD case, in analogy with (1) (Shivamoggi \cite{Shi}), 

\begin{equation}
E(\mathbf{k}) \sim \frac{\mathbf{k}\cdot \mathbf{C_{A}}}{k_\perp^{3}\hat\tau}
\end{equation}

\noindent
$\hat \tau$ being the MHD turn-over time, and $\mathbf{C_{A}}$ being the velocity of Alfv\'{e}n waves in the total magnetic field - 

\begin{equation}\mathbf{
C_{A} = C_{A_{0}} + \tilde{C}_{A}}
\end{equation}

\noindent
where,

\begin{equation}
\mathbf{C_{A_{0}}} \equiv \frac{\mathbf{B_0}}{\sqrt{\rho}}, \mathbf{\tilde{C}_{A}} \equiv \frac{\mathbf{\tilde{B}}}{\sqrt{\rho}}.
\end{equation}

\noindent
$\mathbf{B_0}$ is the applied magnetic field  $\mathbf{\tilde{B}}$ is the magnetic field of the large-scale eddies, and $k_\parallel$ and $k_\perp$ are the wave number components parallel and perpendicular to the magnetic field. $\mathbf{k}\cdot \mathbf{C_A}$ is a measure of the energy involved in the magnetic field-line bending (without which the magnetic field has no participation in the dynamics).

Combining (6) with (3), we obtain\footnote{We assume that MHD turbulence is in a steady state so there is \emph{equi-partition} of kinetic and magnetic energy associated with the advent of an \emph{Alfv\'enic} state ($\mathbf{V}=\pm \mathbf{B}$)(Matthaeus and Montgomery \cite{Mat}).} for the MHD eddy turn-over time $\hat\tau$ - 

\begin{equation}
\hat{\tau} \sim \tau\left(1 + \frac{\tau}{\tau_A}\right)
\end{equation}

\noindent
where,

\begin{equation}
\tau_{A} \sim \frac{1}{k_\parallel C_{A_0}}, \ \ \ \ \tau \sim \frac{1}{k_\perp V} \sim \frac{1}{k_\perp \tilde{C}_{A_\perp}}.
\end{equation}

\noindent
(9) implies that the energy transfer in the MHD case is {\it non-local} in the spectral space which reflects the fact that a large-scale magnetic field cannot be transformed away via Galilean invariance [2]. On the other hand, a mean magnetic field suppresses energy cascade parallel to it via more rapid Alfv\'enic decorrelation in the parallel direction so the energy spectrum would essentially be determined by 2D fluctuations \cite{She}. We have from (9) 

\begin{equation}
\hat{\tau} \sim \left\{
\begin{array}{ll}
\frac{\tau^2}{\tau_A}, & \tau \gg \tau_A \\
\tau, & \tau \ll \tau_A.    
\end{array}
\right. \tag{11a,b}
\end{equation}

\noindent
(11a) corresponds to the case with a very strong applied magnetic field ($B_o \gg$ \~B) and represents the IK hypothesis, so the IK model is pertinent for a strongly magnetized plasma which is indeed anisotropic (invalidating the isotropy assumption in the IK model). (11b) corresponds to the case with a very weak applied magnetic field ($B_o \ll$ \~B) and represents the GS hypothesis. 

\bigskip

(11a) leads to the IK spectrum - 

\begin{equation}
E(k_\perp) \sim \varepsilon^{\frac{1}{2}}C_{A_o}^{\frac{1}{2}}k_\perp^{-\frac{3}{2}} \tag{12}
\end{equation}

\pagebreak
\noindent
while (11b) leads to the GS spectrum - 

\begin{equation}
E(k_\perp) \sim \varepsilon^{\frac{2}{3}}k_\perp^{-\frac{5}{3}}. \tag{13}
\end{equation}

Let us now extend this formulation to include compressibility effects. It may be mentioned that the effects of compressibility on MHD turbulence are not well understood (Lithwick and Goldreich [24], Cho and Lazarian [25]). The following development should therefore be viewed as tentative. Assuming barotropic fluid and adiabatic flow processes, \emph{scale invariance} of compressible Navier-Stokes equations leads to (Shivamoggi\cite{Shi2},\cite{Shi3}) the following scaling behavior of the velocity and density increments - 

\begin{equation}
V(\ell)\sim\ell^\alpha, \rho(\ell)\sim \ell^\frac{2\alpha}{\gamma-1}\tag{14} 
\end{equation}

\noindent
which implies the relation - 

\begin{equation}
\rho\sim \rho_o\left(\frac{V}{c}\right)^{\frac{2}{\gamma-1}}\tag{15}
\end{equation}

\noindent
c being the speed of sound and $\gamma$ the ratio of specific heats of the fluid.

Writing the spectral energy density now as

\begin{equation}
E(k)\sim\frac{\rho V}{k^2 \tau}\tag{16}
\end{equation}

\noindent
the hydrodynamic eddy turn-over time $\tau $, given by (2) then becomes

\begin{equation}
\tau\sim \rho_o^{\frac{\gamma-1}{2\gamma}}c^{-\frac{1}{\gamma}}E^{\frac{1-\gamma}{2\gamma}}k^{\frac{1-3\gamma}{2\gamma}}.
\tag{17}
\end{equation}

In the zero-compressibility limit ($\gamma \rightarrow \infty$), (17), of course, reduces to (3). Using the relation (4), (17) leads to the compressible energy spectrum ([22],[23]) - 

\begin{equation}
E(k)\sim \rho_o^{\frac{\gamma-1}{3\gamma-1}}c^{-\frac{2}{3\gamma - 1}}\varepsilon^{\frac{2\gamma}{3\gamma - 1}}k^{-\frac{5\gamma - 1}{3\gamma - 1}}. \tag{18}
\end{equation}

One may now write for the MHD case, in analogy with (16), 

\begin{equation}
E(\mathbf{k})\sim \frac{\rho\mathbf{k}\cdot\mathbf{C_A}}{k_\perp^3\hat\tau_c}.\tag{19}
\end{equation}

Combining (19) with (15) and (17), we obtain for the compressible MHD eddy turn-over time $\hat\tau_c$ - 

\begin{equation}
\hat\tau_c\sim\tau\left(1+\frac{\tau}{\tau_A}\right) \tag{20}
\end{equation}

\noindent 
which is identical to the incompressible MHD result (9)! Therefore, (9) is apparently \emph{form invariant} with respect to the inclusion of compressibility effects.

On the other hand, (20) implies that the IK spectrum in compressible MHD becomes

\begin{equation}
E(k_\perp)\sim \rho_o^{\frac{\gamma-1}{2\gamma-1}} c^{-\frac{2}{2\gamma - 1}}\varepsilon^{\frac{\gamma}{2\gamma-1}}C_{A_o}^{\frac{\gamma}{2\gamma-1}}k_\perp^{-\frac{3\gamma-1}{2\gamma-1}}\tag{21}
\end{equation}

\noindent
while the GS spectrum in compressible MHD becomes 

\begin{equation}
E(k_\perp)\sim \rho_o^{\frac{\gamma-1}{3\gamma-1}} c^{-\frac{2}{3\gamma - 1}}\varepsilon^{\frac{2\gamma}{3\gamma-1}}k_\perp^{-\frac{5\gamma-1}{3\gamma-1}}.\tag{22}
\end{equation}

(21) and (22) reduce to (12) and (13), respectively, in the zero compressibility limit ($\gamma \rightarrow \infty$).

It may be mentioned that Cho and Lazarian [25] found that the \emph{fast}-mode turbulence, in the low-$\beta$ limit (which corresponds to the IK regime), exhibits the Zakharov-Sagdeev spectrum $E(k)\sim k^{-\frac{3}{2}}$ (Zakharov and Sagdeev [26]). On the other hand, Lithwick and Goldreich [24] conjectured that, in the high-$\beta$ limit (which corresponds to the GS regime), the Alfv\'enic turbulence and \emph{slow}-mode turbulence exhibit the scaling $E(k)\sim k^{-\frac{5}{3}}$.

\pagebreak


\begin{thebibliography}{xxx}
\bibitem{Iro} P.S. Iroshnikov: {\it Sov. Astron.} \textbf{7}, 566, (1964).
\bibitem{Kra} R.H. Kraichnan:  {\it Phys. Fluids} {\bf 8}, 1385, (1965). 
\bibitem{Kol1} A.N. Kolmogorov: {\it Dokl. Akad. Nauk. SSSR} {\bf 30}, 4, (1941).
\bibitem{Mon} D.C. Montgomery: {\it Phys. Scr.} {\bf T2/1}, 83, (1982).
\bibitem{She} J.V. Shebalin, W.H. Matthaeus, and D.C. Montgomery: {\it J. Plasma Phys.} {\bf 29}, 525, (1983).
\bibitem{Sri} S. Sridhar and P. Goldreich: {\it Astrophys. J.} {\bf 432}, 612, (1994). 
\bibitem{Gol1} P. Goldreich and S. Sridhar: {\it Astrophys. J.} {\bf 438}, 763, (1995).
\bibitem{Gol2} P. Goldreich and S. Sridhar: {\it Astrophys. J.} {\bf 485}, 680, (1997). 
\bibitem{Pou} A. Pouquet, U. Frisch and J. Leorat: {\it J. Fluid Mech.} {\bf 77}, 321, (1976).
\bibitem{Mar} J. Maron and P. Goldreich: {\it Astrophys. J.} {\bf 554}, 1175, (2001). 
\bibitem{Mul} W.C. Muller, D. Biskamp, and R. Grappin: {\it Phys. Rev. E.} {\bf 67}, 066302, (2003).
\bibitem{Mul2} W.C. Muller and R. Grappin: {\it Phys. Rev. Lett.} {\bf 95}, 114502, (2005). 
\bibitem{Mul3} W.C. Muller and D. Biskamp: {\it Phys. Rev. Lett.} {\bf 84}, 475, (2000).
\bibitem{Han} N.E.L. Haugen, A. Brandenburg, and W. Dobler: {\it Phys. Rev. E} {\bf 70}, 016308, (2004). 
\bibitem{Fyf}D. Fyfe, D.C. Montgomery, and G. Joyce: {\it J. Plasma Phys.} {\bf 17}, 369, (1977).
\bibitem{Lea} R.J. Leamon, C.W. Smith, N.F. Ness, W.H. Matthaeus, and H.K. Wong: {\it J. Geophys. Res.} {\bf 103}, 4775, (1998).
\bibitem{Gol3} M.L. Goldstein and D.A. Roberts: {\it Phys. Plasmas} {\bf 6}, 4154, (1999). 
\bibitem{Bol} S. Boldyrev: {\it Phys. Rev. Lett.} {\bf 96}, 115002, (2006).
\bibitem{Kol} A.N. Kolmogorov: {\it Dokl. Akad. Nauk. SSSR} {\bf 30}, 301, (1941).
\bibitem{Shi} B.K. Shivamoggi: {\it Ann. Phys.} {\bf 253}, 239, (1997); Erratum in {\it Ann. Phys.} {\bf 312}, 270, (2004).
\bibitem{Mat} W.H. Matthaeus and D. Montgomery: in \emph{Statistical Physics and Chaos in Fusion Plasmas}, Ed. W. Horton and L. Reichl, Wiley (1984).
\bibitem{Shi2} B.K. Shivamoggi: \emph{Phys. Lett. A}, {\bf 166}, 243, (1992).
\bibitem{Shi3} B.K. Shivamoggi: \emph{Ann. Phys.}, {\bf 243} 169, (1995).
\bibitem{Lit} Y. Lithwick and P. Goldreich: \emph{Astrophys. J.} {\bf 562}, 279, (2001).
\bibitem{Cho} J. Cho and A. Lazarian: \emph{Phys. Rev. Lett.} {\bf 88}, 245001, (2002).
\bibitem{Zak} V. Zakharov and R.Z. Sagdeev: \emph{Sov. Phys. Dokl.} {\bf 15}, 439, (1970).

\end{thebibliography}
\end{document}